\documentclass[fleqn,10pt]{wlscirep}
\usepackage[utf8]{inputenc}
\usepackage[T1]{fontenc}
\usepackage{graphicx}
\usepackage{here}
\usepackage{tikz}
\usepackage{amsmath}
\usepackage{braket}
\usepackage{ascmac}
\usepackage{comment}
\usepackage{bm}
\usepackage{color}
\usepackage{geometry}
\usepackage{cite}
\usepackage{ulem}

\title{Measurement-based state preparation of Kerr parametric oscillators}

\author[1, 2]{Yuta Suzuki}
\author[1,3]{Shohei Watabe}
\author[2,4]{Shiro Kawabata}
\author[2,4, $\ast$]{Shumpei Masuda}
\affil[1]{Department of Physics, Faculty of Science Division I, Tokyo University of Science, 1-3 Kagurazaka, Shinjuku-ku, Tokyo 162-8601, Japan}
\affil[2]{Research Center for Emerging Computing Technologies (RCECT), National Institute of Advanced Industrial Science and Technology (AIST), 1-1-1, Umezono, Tsukuba, Ibaraki 305-8568, Japan.}
\affil[3]{College of Engineering, Department of Computer Science and Engineering, Shibaura Institute of Technology, 3-7-5 Toyosu, Koto-ku, Tokyo 135-8548, Japan}
\affil[4]{NEC-AIST Quantum Technology Cooperative Research Laboratory,
National Institute of Advanced Industrial Science and Technology (AIST), Tsukuba, Ibaraki 305-8568, Japan.}

\affil[*]{shumpei.masuda@aist.go.jp}

\begin{abstract}
Kerr parametric oscillators (KPOs) have attracted increasing attention in terms of their application to quantum information processing and quantum simulations. 
The state preparation and measurement of KPOs are typical requirements when they are used as qubits. 
The methods previously proposed for state preparations of KPOs utilize modulation of a pump field or an auxiliary drive field. 
We study the stochastic state preparation of a KPO based on homodyne detection, which does not require modulation of a pump field nor an auxiliary drive field, and thus can exclude unwanted effects of possible imperfection in control of these fields. 
We quantitatively show that the detection data, if averaged over a proper time to decrease the effect of measurement noise, has a strong correlation with the state of the KPO, and therefore can be used to estimate the state of the KPO (stochastic state preparation). 
We examine the success probability of the state estimation taking into account the effect of the measurement noise and bit flips. 
Moreover, the proper range of the averaging time to realize a high success probability is obtained by developing a binomial-coherent-state model, which describes the stochastic dynamics of the KPO under homodyne detection.
\end{abstract}
\begin{document}

\flushbottom
\maketitle
\thispagestyle{empty}

\section*{Introduction}

Kerr parametric oscillators (KPOs)\cite{Milburn1991, Wielinga1993, Goto2016} or Kerr-cat qubits, which are parametric phase-locked oscillators in the single-photon Kerr regime~\cite{Kirchmair2013}, have attracted much attention in terms of their application to quantum information processing\cite{Goto2019} and study of quantum many-body systems\cite{Dykman2018, Rota2019}.
KPOs can be implemented\cite{Goto2019, Meaney2014, Wang2019, Grimm2020} by a superconducting resonator with Kerr-nonlinearity driven by an oscillating pump field in the circuit-QED architecture. 
Two stable coherent states of a KPO in opposite phases can be used as qubit states.
In the KPO, the phase-flip error dominates the bit-flip error because of the robustness of the coherent states against photon loss.
Because of such a biased feature of errors, it is expected that quantum error correction for KPOs can be performed with less overhead than for qubits without such biased noise~\cite{Tuckett2019, Ataides2021}.

Quantum annealing\cite{Goto2016, Nigg2017, Puri2017, Zhao2018, Onodera2020, Goto2020, Kanao2021} and universal quantum computation\cite{Cochrane1999, Goto2016, Puri2017} using KPOs were studied theoretically, and single-qubit operations were demonstrated experimentally\cite{Grimm2020}.
Two-qubit gates preserving the biased feature of errors were proposed~\cite{Puri2020}, and
high error-correction performance by concatenating the XZZX surface code~\cite{Ataides2021} with KPOs~\cite{Darmawan2021} was numerically presented.
Other research subjects on KPOs include fast gate operations and controls~\cite{Zhang2017, Wang2019,Kanao2021-2, Xu2022, Kang2022}, spectroscopy\cite{Yamaji2022,Masuda2021},
\textcolor{black}{tomography~\cite{Wang2019,Grimm2020}}, Boltzmann sampling\cite{Goto2018}, effects of strong pump  field\cite{Masuda2021-2}, quantum phase transitions~\cite{Dykman2018,Rota2019}, quantum chaos~\cite{Milburn1991,Goto2021}, and trajectories~\cite{Bartolo2017,Minganti2016}.

The state preparation and measurement of qubits discussed in this paper are typical requirements in implementation of quantum information processing.
In KPO systems, preparations of predetermined qubit states were studied, using modulation of a pump field~\cite{Goto2016,Puri2017,Masuda2021-2} and an additional drive field~\cite{Yamaji2022}.
In this paper, we study a stochastic state preparation of KPOs based on homodyne detection, which does not require modulation of the pump field nor a drive field in contrast to the conventional methods.
Previously, it was shown that a KPO under homodyne detection is basically in either of two coherent states with opposite phases~\cite{Bartolo2017}, and that the state of a qubit based on a KPO can be measured with homodyne detection, however without crucial analysis on detection data \textcolor{black}{with measurement noise}. 
We quantitatively show that the detection data, if it is averaged over a proper time to decrease the effect of measurement noise, has a strong correlation with the state of the KPO, and therefore can be used to accurately estimate in which coherent state the KPO is (stochastic state preparation).
The success probability of the estimation is examined taking into account the effect of the measurement noise and bit flips.
We obtain the proper range of the averaging time to realize high success probability by using a developed minimum model, which describes the stochastic dynamics of the KPO under homodyne detection.
Moreover, we examine the dependence of the success  probability on the measurement efficiency and the relative phase between the pump field and a local oscillator.

It is known that Rx~\cite{Goto2016-2} and ZZ gates~\cite{Masuda2022} can be performed without modulation of the pump and drive fields. 
Our method of state preparation will be useful in experimental studies of the gate operations, for example aiming at higher fidelity, because the method can exclude unwanted effects of possible imperfection in controls of the pump and drive fields.
Furthermore, our method can offer implementation of the quantum information processing based on KPOs without temporal controls of the pump amplitude, because the universal gate sets~\cite{Goto2016-2} can also be performed without modulation of the pump amplitude.

\section*{Model \textcolor{black}{and methods}}
\begin{figure}[H]
  \centering
  \includegraphics[width=5cm]{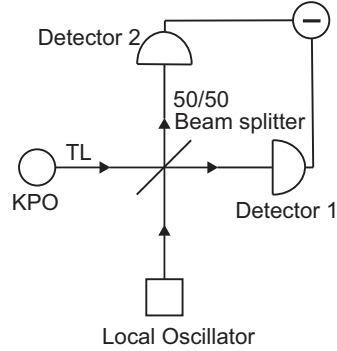}
  \caption{
Schematic illustration of homodyne detection of \textcolor{black}{a} KPO attached to a transmission line (TL). 
The signal from the KPO and the classical coherent light from a local oscillator passing through a beam splitter are detected by detector 1 and 2.
Information about the KPO is obtained after subtraction of the photocurrents at the detector 1 and 2 (Circle with a horizontal line).
}
 \label{Equipment}
\end{figure}

We consider homodyne detection of a KPO illustrated in Fig.~\ref{Equipment}.
Classical coherent light generated by a local oscillator and microwave photons emitted from the KPO are splitted by a 50/50 beam splitter and are detected at detectors 1 and 2.
The KPO is connected to a transmission line (TL), where the emitted photons propagate.

In order to take into account the effect of the homodyne detection on the density matrix of the KPO $\rho_{\rm c}$, we use a stochastic master equation (SME)  represented as~\cite{Bartolo2017, Wiseman2009}
\begin{equation}
\begin{split}
\rho_{\rm{c}}(t+\tau)
&=
\rho_{\rm{c}}(t)
-i
\left[
-
\frac{\chi}{2}\hat{a}^\dagger \hat{a}^\dagger \hat{a}\hat{a}
+
\beta(\hat{a}^\dagger \hat{a}^\dagger+\hat{a}\hat{a})
,\rho_{\rm{c}}(t)\right]\tau
+\left[
\kappa\hat{a}\rho_{\rm{c}}(t)\hat{a}^{\dagger}
-
\frac{\kappa}{2}\left\{\hat{a}^\dagger\hat{a},\rho_{\rm{c}}(t)\right\}
\right]
\tau
\\
&\quad
-i\sqrt{\kappa}
\left[
\exp \left(-i\Theta_{\rm{LO}}\right)\hat{a}\rho_{\rm{c}}(t)
-
\exp \left(i\Theta_{\rm{LO}}\right)\rho_{\rm{c}}(t)\hat{a}^{\dagger}
\right]\Delta W(t)
-\mathrm{Tr}[\rho_{\rm{c}}(t)\hat{A}_{\Theta_{\rm{LO}}}]
\rho_{\rm{c}}(t)
\Delta W(t),
\\
\end{split}
\label{stochastic master eq}
\end{equation}
where $\chi$,  $\beta$ and $\kappa$ are the anharmonicity parameter of the KPO, amplitude of the pump field and the decay rate to the TL, respectively.
We refer readers to, e.g., Refs.~[\citenum{Wang2019,Masuda2021-2}] for the connection between the system parameters to circuit models of KPOs. 
In Eq.~(\ref{stochastic master eq}), $\Theta_{\rm{LO}}$ is the relative phase of the classical coherent light of the local oscillator and the pump field.
$\Delta W$ is the noise in the photon numbers measured by the two detectors, which is assumed to be Gaussian white noise with the mean of $0$ and variance $\tau$, and $\Delta W^2=\tau$ ~\cite{Wiseman2009}.
We hereafter refer to $\Delta W$ as noise. 
$\hat{a}$ is the annihilation operator for the KPO, and $\hat{A}_{\Theta_{\rm{LO}}}$ is defined by $\hat{A}_{\Theta_{\rm{LO}}}
= i\sqrt{\kappa}[
\exp\left(i\Theta_{\rm{LO}}\right)\hat{a}^\dagger - \exp\left(-i\Theta_{\rm{LO}}\right)\hat{a}]$.
The solution $\rho_{\rm c}(t)$ of the SME~(\ref{stochastic master eq}) for a given $\Delta W$ represents one possible realization of the dynamics under homodyne detection.
The ensemble average of $\rho_{\rm{c}}(t)$ over $\Delta W$ in the SME~(\ref{stochastic master eq}) coincides with the density operator of the master equation~(S2) which governs time evolution of the KPO when it is not measured (See Supplementary Section S1 for the Hamiltonian and master equation of a KPO).

For $\kappa/4|\chi \alpha|^2 \ll 1$ , which \textcolor{black}{was} realized, e.g. in Ref.[\citenum{Wang2019}], the stationary state of the master equation~(S2) is approximately represented as $(\ket{\alpha}\bra{\alpha}+\ket{-\alpha}\bra{-\alpha})/2$ with~\cite{Meaney2014, Puri2017-2}
\begin{equation}
\begin{split}
|\alpha|=\left(\frac{4\beta^2-\kappa^2/4}{\chi^2}\right)^{1/4}
,~~\
\rm{arg}[\alpha]=\frac{1}{2}\sin^{-1}\left(-\frac{\kappa}{4\beta}\right).
\end{split}
\label{Puri}
\end{equation}
\noindent
As shown later, the state of the KPO jumps between $\ket{\alpha}$ and $\ket{-\alpha}$.
We aim at stochastically preparing either of the states using the data measured by the detectors. 
In order to evaluate the efficiency of the protocol, we use the fidelities defined by
$F_{\pm}=\mathcal{F}[\rho_{\rm{c}}(t),\ket{\pm\alpha}\bra{\pm\alpha}]$, where $\mathcal{F}[\rho_{\rm a},\rho_{\rm b}]=\left(\mathrm{Tr}[\sqrt{\sqrt{\rho_{\rm a}}\rho_{\rm b} \sqrt{\rho_{\rm a}}}]\right)^2$~\cite{Nielsen2010}. 
(The fidelities between the state of a KPO and these coherent states have not been examined with SME (\ref{stochastic master eq}) to the best of our knowledge.)
\textcolor{black}{In numerical simulations, w}e assume the followings:  
the initial state of the KPO is $(\ket{\alpha}\bra{\alpha}+\ket{-\alpha}\bra{-\alpha})/2$ to which the KPO relaxes due to the decay to the TL~\cite{Puri2017-2} when the KPO is not measured; 
all the system parameters are fixed during the homodyne detection. 
\textcolor{black}{We used QuTiP~\cite{QuTiP} for a part of numerical simulations.}

\section*{Results}
\begin{figure}[H]
  \centering
  \includegraphics[width=15cm]{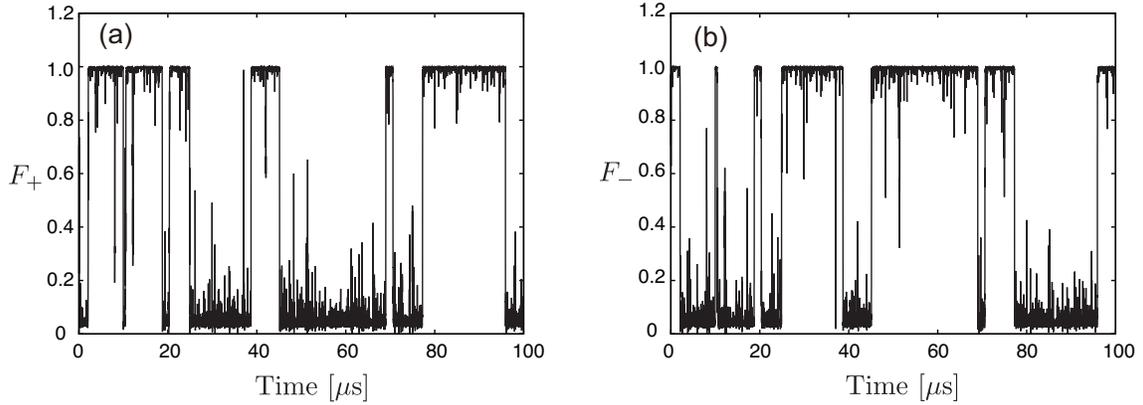}
  \caption{
Time dependence of the fidelities, $F_{\pm}=\mathcal{F}[\rho_{\rm{c}}(t),\ket{\pm\alpha}\bra{\pm\alpha}]$. Panels (a) and (b) are for  $F_{+}$ and $F_{-}$, respectively.
The used parameters are $\chi/2\pi=3~\rm{MHz}$, $\beta/2\pi=3~\rm{MHz}$ and $\kappa/2\pi=3~\rm{MHz}$.
$\alpha$ given by Eq.~(\ref{Puri}) is approximately $1.38-0.18i$. 
These parameters are experimentally feasible~\cite{Wang2019}.
}
\label{FidPM}
\end{figure}

Figure~\ref{FidPM}(a,b) shows the time dependence of the fidelities $F_{\rm \pm}$,  
where we assumed that there is no photon loss (the effect of photon loss is examined in the section entitled ``Imperfect detection").
The time dependence of $F_{\rm \pm}$ implies that the state of the KPO jumps between $\ket{\alpha}$ and $\ket{-\alpha}$, and remains in either of the coherent states between jumps.
Importantly, we cannot obtain $F_{\rm \pm}$ in actual measurements.
In the following, we investigate how accurately we can estimate the state of the KPO from the measurement results.

\subsection*{State estimation}
Measurement results that observers can obtain in the homodyne detection is the difference between the numbers of photons detected by the two detectors.
We use this data for the estimation of the state of the KPO.
The difference between the numbers of photons detected by detectors 1 and 2 from $t$ to $t+\tau$ is represented as \cite{Gambetta2008,Wiseman2009}
\begin{equation}
\begin{split}
\Delta N(t)=\frac{1}{\varepsilon}
\left(
\Delta W(t)+
\tau\mathrm{Tr}[\rho_{\rm{c}}\hat{A}_{\Theta_{\rm{LO}}}]
\right),
\end{split}
\label{dif photon num}
\end{equation}
where $\tau$ is much smaller than $\beta^{-1}, \chi^{-1}$ and $\kappa^{-1}$.
Here, $\epsilon^{-1}$ is the product of the square root of phase velocity in the TL and the intensity of the classical coherent light \cite{Wiseman2009}.
When the KPO is in either the two coherent states, that is, $\rho_c=\ket{\pm\alpha}\bra{\pm\alpha}$,
$\Delta N$ can be written as
\begin{equation}
\begin{split}
\Delta N_\pm=
\frac{1}{\epsilon}
\left(
\Delta W(t)\pm
2|\alpha|\sqrt{\kappa}\tau
\sin(\delta \theta)
\right)
\end{split}
\label{dif photon num +}
\end{equation}
with $\delta \theta=\mathrm{arg}[\alpha]-\Theta_{\mathrm{LO}}$.
Importantly, the sign and amplitude of the second term depend on the state of the KPO and $\delta\theta$, respectively.
We mainly discuss the case for $\delta \theta=\pi/2$, which maximizes the second term of Eq.~(\ref{dif photon num +}).
The effect of the deviation of $\delta \theta$ from $\pi/2$ is examined in the section entitled ``Imperfect detection".

If the amplitude of the noise $|\Delta W|$ is always smaller than $2|\alpha|\sqrt{\kappa}\tau$, we can identify the state of the KPO from the sign of $\Delta N(t)$.
However, as shown below, $|\Delta W|$ can be larger than $2|\alpha|\sqrt{\kappa}\tau$.
Therefore, it is important to take time average of $\Delta N(t)$ for a certain period of time to decrease the effect of the noise. 
The photon-number difference averaged from $t-T_{\rm{a}}$ to $t$ is represented as
\begin{equation}
\bar{N}(t,T_{\rm a})=\frac{\tau}{T_{\rm a}} \sum_{k=0}^{T_{\rm a}/\tau}\Delta N (t-k\tau),
\label{dif photon num ave}
\end{equation}
where we assume $T_{\rm a}$ is integer multiple of $\tau$.

\begin{figure}[H]
  \centering
  \includegraphics[width=15cm]{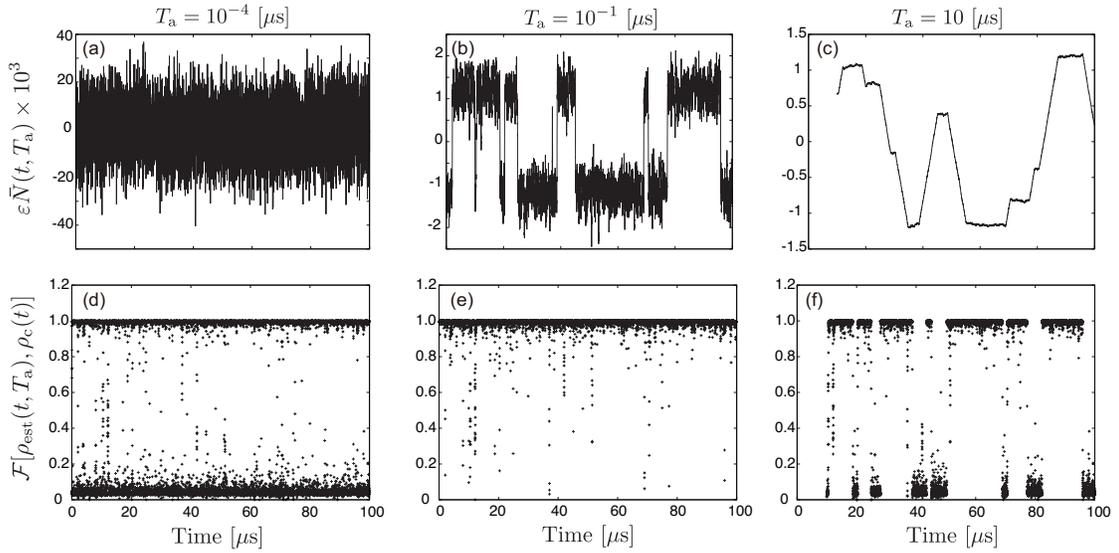}
  \caption{
Time dependence of $\bar{N}$ and $\mathcal{F}[\rho_{\rm{est}}(t,T_{\rm a}),\rho_{\rm{c}}(t)]$ for $T_{\rm{a}}=10^{-4}~\rm{\mu s}$ (a,d), $T_{\rm{a}}=10^{-1}~\rm{\mu s}$ (b,e), $T_{\rm{a}}=10~\rm{\mu s}$ (c,f).
The used parameters are the same as in Fig.~\ref{FidPM}.
The data is represented by lines and dots in the upper and lower panels, respectively.}
\label{ALL_3_30_22}
\end{figure}

We estimate the state of the KPO \textcolor{black}{at time $t$} using the sign of $\bar{N}$, that is, we estimate the KPO to be in $\ket{\alpha}\bra{\alpha}$ for $\bar{N}>0$ and $\ket{-\alpha}\bra{-\alpha}$ for $\bar{N}<0$, respectively.
The estimated state is represented as
\begin{equation}
\rho_{\rm{est}}(t,T_{\rm a})
=
\left\{
\begin{array}{ll}
\ket{\alpha}\bra{\alpha} & (\bar{N}(t,T_{\rm a})>0),\\
\ket{-\alpha}\bra{-\alpha} & (\bar{N}(t,T_{\rm a})<0).
\end{array}
\right.
\label{rho_est}
\end{equation}
Figure~\ref{ALL_3_30_22}(a-c) shows the time dependence of $\bar{N}$ for various values of $T_{\rm a}$.
For $T_{\rm{a}}=10^{-4}$~$\mu$s, the fluctuation of $\bar{N}$ is too larger to identify the state of the KPO due to the noise (Fig.~\ref{ALL_3_30_22}(a)).
On the other hand, for $T_{\rm{a}}=10^{-1}$~$\mu$s, $\bar{N}$ approximately takes either of $\pm2|\alpha|\sqrt{\kappa}\tau$ (Fig.~\ref{ALL_3_30_22}(b)).
For $T_{\rm{a}}=10$~$\mu$s, the second term of Eq.~(\ref{dif photon num +}) is smeared because of bit flips and the long averaging time (Fig.~\ref{ALL_3_30_22}(c)). 

Figure~\ref{ALL_3_30_22}(d-f) shows the fidelity of the estimation defined by $\mathcal{F}[\rho_{\rm{est}}(t,T_{\rm a}),\rho_{\rm{c}}(t)]$. 
The fidelity is close to 0 or 1 most of the time, and thus the distribution of the fidelity is bimodal. 
The fidelity of approximately zero corresponds \textcolor{black}{to the case that the estimated state is $\ket{\pm\alpha}$ while the KPO is actually in $\ket{\mp\alpha}$}. 
The fidelity is larger than 0.99 most of the time for $T_{\rm{a}}=10^{-1}$~$\rm{\mu s}$.
On the other hand, the fidelity for $T_{\rm{a}}=10^{-4}$~$\mu$s and $10$~$\mu$s can become approximately zero due to the too short and too long averaging times, respectively, thus the time-averaged fidelity is decreased.

\subsection*{Proper averaging time for accurate estimation}
\begin{figure}[H]
  \centering
  \includegraphics[width=8cm]{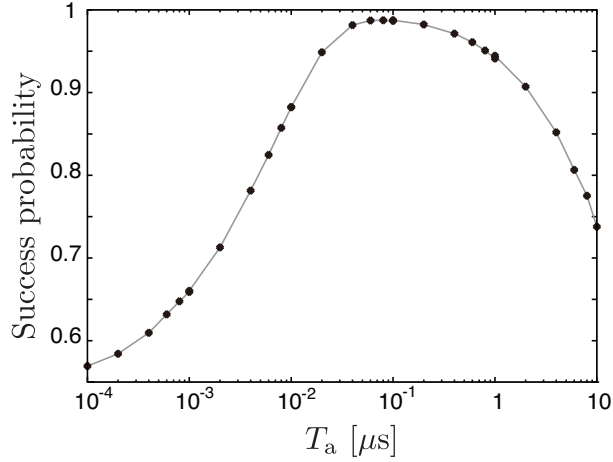}
  \caption{
Success probability, defined by $\mathcal{F}[\rho_{\rm{est}}(t,T_{\rm a}),\rho_{\rm{c}}(t)]$ time-averaged over a 1000~$\rm \mu$s period, as a function of $T_{\rm a}$. 
The used parameters are the same as in Fig.~\ref{FidPM}.
  }
\label{isousa_kotei_error_Time}
\end{figure}

Figure~\ref{isousa_kotei_error_Time} shows the $T_{\rm a}$ dependence of $\mathcal{F}[\rho_{\rm{est}}(t,T_{\rm a}),\rho_{\rm{c}}(t)]$ time averaged over a $1000~\mu$s period.
Hereafter, we refer to the averaged fidelity as the success probability of estimation.
The success probability is higher than 0.987 around $T_{\rm a}=10^{-1}$~$\rm{\mu}$s.
It is clearly seen that there is a proper range of $T_{\rm a}$ to obtain the high success probability.
The proper range of $T_{\rm a}$ for a given value of the success probability $K$ is bounded from below due to the noise and bounded from above due to smearing by the time averaging.
In the following, we obtain the upper bound $T_{{K}}^{\rm{U}}$ and lower bound $T_{{K}}^{\rm{L}}$ of $T_{\rm{a}}$ for a given success probability $K$.

\subsubsection*{Lower bound $T_{K}^{\rm{L}}$}
We consider the case that the averaging time $T_{\rm a}$ is much shorter than the typical duration in which the KPO remains in either of $\ket{\pm\alpha}$. 
Then, $\Delta {N}(t)$ is typically represented as Eq.~(\ref{dif photon num +}) and fluctuates around either of 
$\pm 2|\alpha|\sqrt{\kappa}\tau/\varepsilon$ due to the Gaussian noise $\Delta W$, except when jumps occur.
The fluctuation of $\bar{N}(t)$ has the Gaussian distribution with the standard deviation of $\sigma(T_{\rm a}) = \sqrt{{\tau^2}/{T_{\rm a}\varepsilon^2}}$, where the effect of jumps to $\bar{N}(t)$ is neglected because jumps seldom occur in $T_{\rm a}$.
Then, the success probability $K$ can be related to $T_{\rm a}$ as 
\begin{equation}
K = \int_{-2|\alpha|\sqrt{\kappa}\tau/\varepsilon}^{\infty} dx \frac{1}{\sigma(T_{\rm a})\sqrt{2\pi}} \exp\Big{[}\frac{x^2}{2\sigma^2(T_{\rm{a}})}\Big{]},
\label{K_4_2_22}
\end{equation}
where $K$ is the same as the ratio of the colored area to the total area under the Gaussian curve illustrated in Fig.~\ref{bound}(a). 
$T_{\rm a}$ in Eq~(\ref{K_4_2_22}) equals to the lower bound of the averaging time, $T_{K}^{\rm{L}}$, to obtain the success probability higher than or equal to $K$. 
Therefore, we can obtain $T_{K}^{\rm{L}}$ by solving Eq.~(\ref{K_4_2_22}). 
For example, $T_{K}^{\rm{L}}$ for $K=0.95$ is
\begin{equation}
\begin{split}
T_{0.95}^{\rm{L}}=\frac{1.65^2}{4|\alpha|^2 \kappa}
.
\end{split}
\label{95L}
\end{equation}

\subsubsection*{Upper bound $T_{K}^{\rm{U}}$}
Averaging over a long period of time can degrade the accuracy of the estimation of the state of the KPO due to the jump.
We assume that this smearing effect determines $T_{K}^{\rm{U}}$, and derive $T_{K}^{\rm{U}}$ by developing a binomial-coherent-state model that describes the stochastic dynamics of the KPO in Eq.~(\ref{stochastic master eq}).

\begin{figure}[H]
  \centering
  \includegraphics[width=11cm]{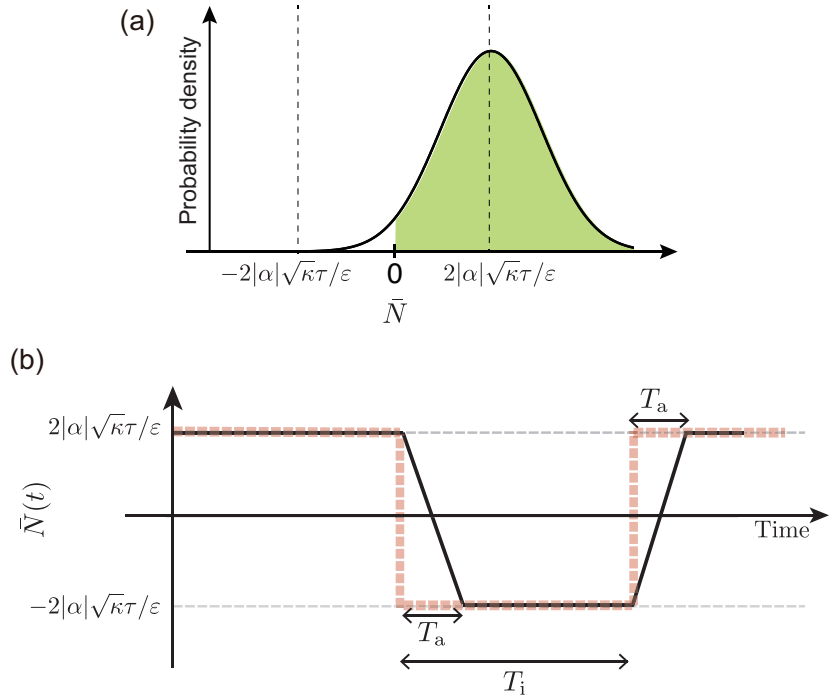}
  \caption{
Panel(a): Schematic illustration of the distribution of $\bar{N}$ fluctuating around $\Delta N'=2|\alpha|\sqrt{\kappa}\tau/\varepsilon$.
The ratio of the colored area to the total area under the Gaussian curve is the same as the success probability $K$.
Panel(b): Schematic illustration of a typical time evolution of $\bar{N}$ in the binomial-coherent-state model in which the noise is neglected.
The dashed and solid lines represent $\bar{N}$ for $T_{\rm{a}}=0$ (without time averaging) and $0<T_{\rm{a}}<T_{\rm{i}}$ (with time averaging), respectively.
  }
  \label{bound}
\end{figure}

In the binomial-coherent-state model, we assume that the state of the KPO can only take either of $\ket{\pm\alpha}$, and jumps between them with a probability of $p~(=\Omega dt)$ in a small time $dt$. 
This stochastic process can be represented as the binomial process in the two coherent states. 
Then, the mean time interval between jumps is 
\begin{eqnarray}
E[T_{\rm{i}}]=\frac{1}{\Omega},
\label{ET_4_4_22}
\end{eqnarray}
\noindent
because $\Omega$ is the average rate of jumps.
Figure~\ref{bound}(b) illustrates a typical time evolution of $\bar{N}$.
Due to smearing effect and jumps, wrong estimations occur in the period of $T_{\rm{a}}/2$ per jump.
The error rate, defined by the ratio of the duration of the wrong estimation to the total measurement time, can be written as $T_{\rm{a}}/2E[T_{\rm{i}}]$, and the error rate is also written as $1-K$.
Thus, we obtain $1-K=T_{K}^{\rm{U}}/2E[T_{\rm{i}}]$, where we replaced $T_{\rm{a}}$ by $T_{K}^{\rm{U}}$.
Using Eq.~(\ref{ET_4_4_22}), we obtain $T_{K}^{\rm{U}}$ as 
\begin{eqnarray}
T_{K}^{\rm{U}} =2(1-K)/\Omega.
\label{TKU_4_4_22}
\end{eqnarray}

Now, we obtain $\Omega$ by the following manner.
In the binomial-coherent-state model, the ensemble average of the density operator can be represented as
\begin{equation}
\bar{\rho}_{\rm b}(t)=
\sum_{k=2n}^{N}{}_N \mathrm{C}_k p^{k} (1-p)^{N-k} \ket{\alpha}\bra{\alpha} 
+
\sum_{k=2n+1}^{N}{}_N \mathrm{C}_k p^{k} (1-p)^{N-k} \ket{-\alpha}\bra{-\alpha} 
,
\label{rho_4_5_22}
\end{equation}
where $N=t/dt$, and we assumed that the KPO is in $\ket{\alpha}$ at the initial time.
As shown in Supplementary Section S2, the expectation value of $\hat{x}=(\hat{a}+\hat{a}^\dagger)/2$ corresponding to $\bar{\rho}_{\rm b}(t)$ in Eq.~(\ref{rho_4_5_22}) is written as $\langle \hat{x} \rangle = {\rm Re}[\alpha] \exp(-2\Omega t)$ in the limit of $dt\rightarrow 0$.
Because the binomial-coherent-state model approximates the dynamics governed by the SME, $\bar{\rho}_{\rm b}(t)$ approximately coincides with the solution of the master equation~(S2) (Note that the ensemble average of $\rho_{\rm{c}}(t)$ over $\Delta W$ coincides with the density operator of the master equation).
Therefore, we can obtain $\Omega$ by fitting ${\rm Re}[\alpha] \exp(-2\Omega t)$ to $\braket{\hat{x}}$ with the master equation~(S2) (the detailed discussion and results of the fitting can be found in Supplementary Section S2).

\subsubsection*{Numerical results}
\begin{figure}[H]
  \centering
  \includegraphics[width=15cm]{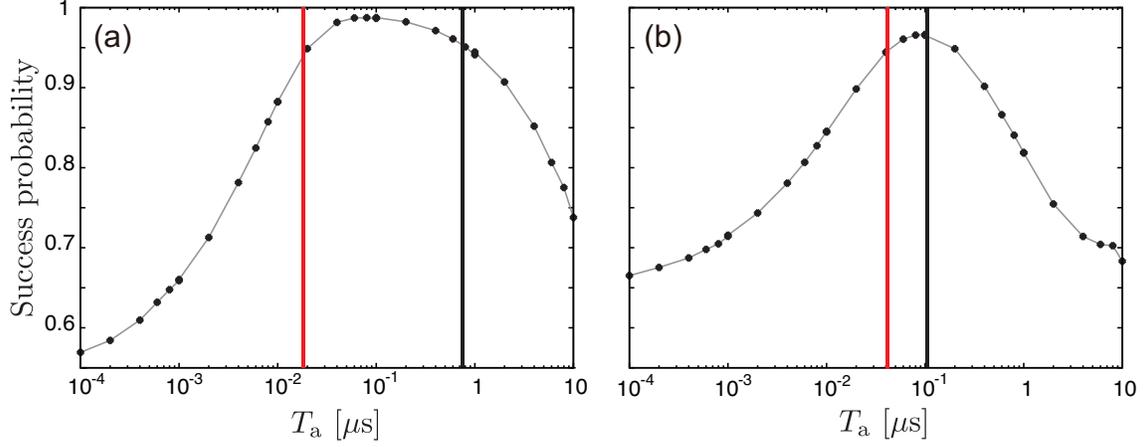}
  \caption{
$T_{\rm a}$ dependence of the mean of $\mathcal{F}[\rho_{\rm{est}}(t,T_{\rm a}),\rho_{\rm{c}}(t)]$ for
$\kappa=\chi$, $\beta=\chi$ (a) and $\kappa=\chi$, $\beta=\chi/2$ (b).
The parameters for panels (a) and (b) correspond $\alpha=1.38-0.18i$ and $\alpha=0.90-0.24i$, respectively.
The vertical red (black) line represents $T_{0.95}^{\rm{L(U)}} = 1.86\times10^{-2}~\rm{\mu s}~(7.52\times10^{-1}~\rm{\mu s})$ in panel (a) and $T_{0.95}^{\rm{L(U)}} =4.17\times10^{-2}~\rm{\mu s}~(1.04\times10^{-1}~\rm{\mu s})$ in panel (b).
The other parameters are the same as in Fig.~\ref{FidPM}.
}
\label{TU-ALL_4_7_22}
\end{figure}

Figure~\ref{TU-ALL_4_7_22} shows the $T_{\rm a}$ dependence of the success probability together with $T_{0.95}^{\rm{L(U)}}$ for two different parameter sets.
It is seen that the values of $T_{\rm{K}}^{\rm{L(U)}}$ obtained in the above section approximate well the numerical ones.

\begin{figure}[H]
  \centering
  \includegraphics[width=7cm]{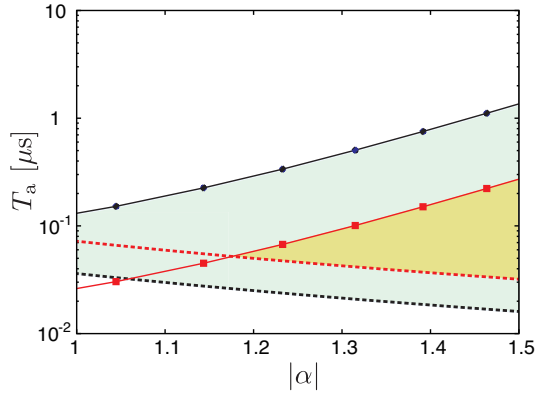}
  \caption{
$T_{K}^{\rm U}$ and $T_{K}^{\rm L}$ as a function of $|\alpha|$ in Eq.~(\ref{Puri}).
The dots and dashed curves are for $T_{K}^{\rm U}$ and $T_{K}^{\rm L}$, respectively.
The solid curves for $T_{K}^{\rm U}$ are guide to the eye.
The black and red data are for $K=0.95$ and 0.99, respectively, where $\beta$ was changed, while $K$ is fixed, in order to change $|\alpha|$.
The other parameters are the same as in Fig.~\ref{FidPM}.
}
  \label{Ta_range}
\end{figure}

Figure~\ref{Ta_range} represents the high success probability regime in the $T_{\rm a}$-$|\alpha|$ plane.
It is seen that $T_K^{\rm L}$ decreases with the increase of $|\alpha|$ as analytically exemplified in Eq.~(\ref{95L}) because the effect of the noise to the result of the estimation becomes small for large $|\alpha|$. On the other hand, $T_K^{\rm U}$ increases with $|\alpha|$ because $E[T_{\rm i}]$ increases exponentially with $|\alpha|^2$~~\cite{Puri2019} as shown in Supplementary Section S2. Thus, the range of $T_{\rm a}$, which gives the high success probability, increases with $|\alpha|$. The maximum success probability also increases with $|\alpha|$. 
We attribute this to the fact that the two quasi stable states, between which the KPO jumps, can be approximated by $\ket{\pm \alpha}$ more accurately in Eq.~(\ref{Puri}) when $|\alpha|$ increases~\cite{Puri2020}.

\subsection*{Imperfect detection}
In the previous sections, we considered the ideal homodyne detection without photon loss, and we set $\delta\theta=\pi/2$ in order to maximize the amplitude of the second term of Eq.~(\ref{dif photon num +}).
In this section, we examine the effect of the photon loss and the deviation of $\delta\theta$ from the ideal value on the success probability of estimation. 

We consider the case that a proportion $\eta$ of photons are detected, while the rest are lost. 
We refer $\eta$ as the efficiency of the measurement.
For the measurement with the efficiency $\eta$, the SME and measurement result are represented as~\cite{Jacob2006, Wiseman2009}
\begin{equation}
\begin{split}
\rho_{c}(t+\tau)&=
\rho_{c}(t)
-i
\left[
\left(\omega_{s}-\chi-\frac{\omega_p}{2}\right)\hat{a}^\dagger \hat{a}
-
\frac{\chi}{2}\hat{a}^\dagger \hat{a}^\dagger \hat{a}\hat{a}
+
\beta(\hat{a}^\dagger \hat{a}^\dagger+\hat{a}\hat{a})
,\rho_{c}(t)\right]\tau
\\
&\quad
+\left[
\kappa_{ex}\hat{a}\rho_{c}(t)\hat{a}^{\dagger}
-
\frac{\kappa_{ex}}{2}\left\{\hat{a}^\dagger\hat{a},\rho_{c}(t)\right\}
\right]
\tau
-i\sqrt{\kappa_{ex}}
\left[
\exp \left(-i\Theta_{LO}\right)\hat{a}\rho_{c}(t)
-
\exp \left(i\Theta_{LO}\right)\rho_{c}(t)\hat{a}^{\dagger}
\right]\sqrt{\eta}\Delta W(t)
\\
&\quad
-\mathrm{Tr}[\rho_{c}(t)\hat{A}_{\Theta_{LO}}]
\rho_{c}(t)
\sqrt{\eta}\Delta W(t), 
\end{split}
\label{SME eta theta}
\end{equation}
and
\begin{equation}
\begin{split}
\Delta N(t)=\frac{1}{\varepsilon}
\left(
\sqrt{\eta}\Delta W+
\eta \tau\mathrm{Tr}[\rho_{\rm{c}}\hat{A}_{\Theta_{LO}}]
\right).
\end{split}
\label{output eta theta}
\end{equation}
As in the previous section, we assume that the averaging time $T_{\rm a}$ is much shorter than the typical duration that the KPO remains in either of $\ket{\pm\alpha}$. 
When $\rho_c=\ket{\pm\alpha}\bra{\pm\alpha}$,
$\Delta N$ fluctuates around $\pm2|\alpha|\sqrt{\kappa}\tau\sin(\delta \theta)\eta/\varepsilon$; the standard deviation of $\bar{N}$ is $\sigma(T_{\rm{a}},\eta)=\sqrt{{\tau^2\eta}/{T_{\rm{a}}\varepsilon^2}}$.
The success probability $K$ can be related to $T_{\rm a}$ as
\begin{equation}
K = \int_{-2|\alpha|\sqrt{\kappa}\tau \sin(\delta \theta) \eta/\varepsilon}^{\infty} dx \frac{1}{\sigma(T_{\rm a},\eta)\sqrt{2\pi}} \exp\Big{[}\frac{x^2}{2\sigma^2(T_{\rm{a}},\eta)}\Big{]}.
\label{K_5_5_22}
\end{equation}
We can obtain $T_{K}^{\rm{L}}$ by solving Eq.~(\ref{K_5_5_22}). 
For example, $T_{K}^{\rm{L}}$ for $K=0.95$ is
\begin{equation}
\begin{split}
T_{0.95}^{\rm{L}}=\frac{1.65^2}{4|\alpha|^2 \kappa\sin^2(\delta \theta)\eta}
.
\end{split}
\label{95L eta}
\end{equation}
On the other hand, $T_{K}^{\rm{U}}$ in Eq.~(\ref{TKU_4_4_22}) does not depend on $\eta$ and $\delta \theta$ because it is derived by using the master equation~(S2) that does not have $\eta$ and $\delta \theta$.

\begin{figure}[H]
  \centering
  \includegraphics[width=15cm]{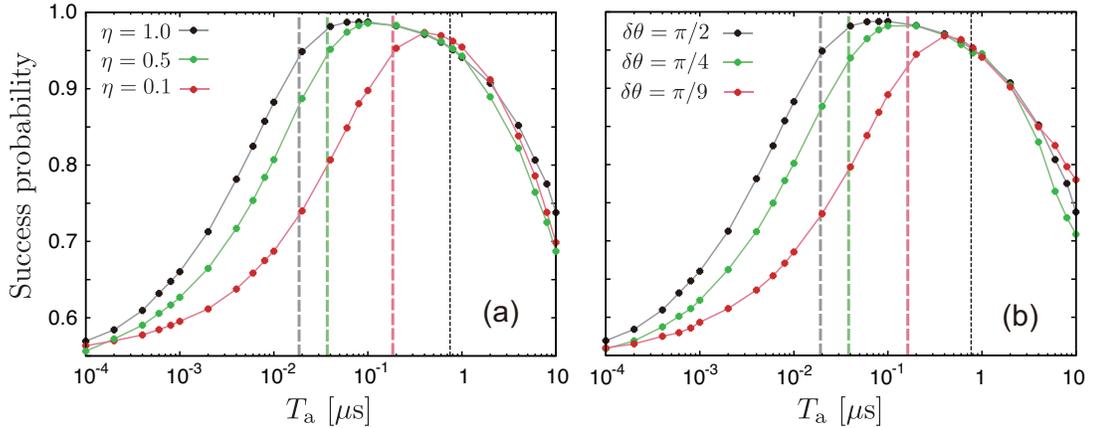}
  \caption{
Dependence of the success probability on $T_{\rm a}$ for various values of $\eta$ and $\delta \theta$.
Panel (a) is for $\delta \theta=\pi/2$; panel (b) is for $\eta=1$.
The other parameters are the same as in Fig.~\ref{FidPM}.
In panel (a), the vertical gray, green and red dashed lines represent $T_{0.95}^{\rm L}=1.86\times10^{-2}$, $3.73\times10^{-2}$ and $1.86\times10^{-1}~\rm{\mu s}$, respectively; in panel (b), the vertical gray, green and red dashed lines represent $T_{0.95}^{\rm L}=1.86\times10^{-2}$, $3.73\times10^{-2}$ and $1.59\times10^{-1}~\rm{\mu s}$, respectively.
The vertical black dotted line is for $T_{0.95}^{\rm U}=7.52\times10^{-1}~\rm{\mu s}$ in both panels.
}
 \label{eta_theta}
\end{figure}

Figure~\ref{eta_theta} shows the dependence of the success probability on $T_{\rm a}$ for various values of $\eta$ and $\delta \theta$ with $T_{K}^{\rm{L}}$ and $T_{K}^{\rm{U}}$.
The success probability decreases on the left side of its peak as $\eta$ decreases or $\delta \theta$ deviates from $\pi/2$. 
On the other hand, the right side of the peak is not sensitive to $\eta$ and $\delta \theta$.
These results are consistent with the analysis of $T_{K}^{\rm{L}}$ and $T_{K}^{\rm{U}}$.

\section*{Conclusions and discussions}
We have studied the stochastic state preparation of a KPO based on homodyne detection.
We have shown that the measured data, if it is time averaged with a proper averaging time to decrease the effect of noise, has a strong correlation with the state of the KPO, and therefore can be used for  estimation of the state of the KPO.
We have quantitatively examined the success probability of the estimation taking into account the effect of the noise and bit flips, and have shown that the success probability is higher than 0.98 with the parameter used.
We have developed a binomial-coherent-state model, which describes the stochastic dynamics of the KPO under homodyne detection, and by using it we have obtained a proper range of the averaging time to realize the high success probability.
Our analysis based on the binomial-coherent-state model implies that the success probability is further increased as the size of the coherent state becomes large.
Furthermore, we have examined the effect of the imperfection of the measurement and the choice of the phase of the coherent light of the local oscillator, on the state estimation.
Our scheme of state preparation of KPOs does not require a drive field nor modulation of the pump field in contrast to conventional methods.  

Although we focused on preparation of the two stable coherent states in this paper, this method followed by single-qubit gates conditioned on measurement results can generate an arbitrary qubit state. 
Our scheme of state preparation can be applied straightforwardly to multi-KPO systems when the time interval of jumps of KPOs is sufficiently long. 
It is also expected that turning on the ferromagnetic or anti-ferromagnetic coupling between KPOs~\cite{Masuda2022} can increase the efficiency of state preparation by mitigating bit flips of individual KPOs.


\section*{Acknowledgements}
The authors thank T. Nikuni, H. Goto, T. Kanao, Y. Matsuzaki, T. Ishikawa, T. Yamaji, A. Yamaguchi and T. Yamamoto for fruitful discussions.
This paper is partly based on results obtained from a project, JPNP16007, commissioned by the New Energy and Industrial Technology Development  Organization (NEDO), Japan.

\section*{Author contributions}
Y.S. carried out the theoretical analysis and numerical simulations. S.W. and S.K. contributed to theoretical analysis. S.M. provided the initial ideas and supervised the work in all respects. The manuscript was written by Y.S. and S.M. with input from the other authors. All authors reviewed the manuscript.


\clearpage
\setcounter{figure}{0}
\renewcommand*{\thefigure}{S\arabic{figure}}
\setcounter{table}{0}
\renewcommand*{\thetable}{S\arabic{table}}
\setcounter{equation}{0}
\renewcommand*{\theequation}{S\arabic{equation}}

\begin{center}
{\LARGE \bf Supplemental information:\\
Measurement-based state preparation of Kerr parametric oscillators}\\ \vspace{0.4cm}

{\Large Yuta Suzuki$^{1,2}$, Shohei Watabe$^{1,3}$, Shiro Kawabata$^{2,4}$ and\\ Shumpei Masuda$^{2,4,\ast}$}\\
\vspace{0.4cm}
$^{1}$ Department of Physics, Faculty of Science Division I, Tokyo University of Science, 1-3 Kagurazaka, Shinjuku-ku, Tokyo 162-8601, Japan.\\
$^{2}$ Research Center for Emerging Computing Technologies (RCECT), National Institute of Advanced Industrial Science and Technology (AIST), 1-1-1, Umezono, Tsukuba, Ibaraki 305-8568, Japan.\\
$^{3}$ College of Engineering, Department of Computer Science and Engineering, Shibaura Institute of Technology, 3-7-5 Toyosu, Koto-ku, Tokyo 135-8548, Japan\\
$^{4}$ NEC-AIST Quantum Technology Cooperative Research Laboratory,
National Institute of Advanced Industrial Science and Technology (AIST), Tsukuba, Ibaraki 305-8568, Japan.

\vspace{0.4cm}
$^\ast$ shumpei.masuda@aist.go.jp
\end{center}

\section*{S1 Hamiltonian and master equation}
\label{Master equation}
The Hamiltonian for the composite system of a Kerr parametric oscillator (KPO) and a transmission line (TL) can be written as
\begin{equation}
\begin{split}
\hat{H}_{\rm{tot}}
&=
\hbar\omega_{s}\hat{a}^\dagger \hat{a}
-
\frac{\hbar\chi}{12}(\hat{a}^\dagger+\hat{a})^4
+
2\hbar\beta(\hat{a}^\dagger+\hat{a})^2 \cos(\omega_p t)
+
\hbar\int_{0}^{\infty}dk v_b k\hat{b}^\dagger_{k} \hat{b}_{k}
+
\hbar \sqrt{\frac{v_b \kappa}{2\pi}}\int_{0}^{\infty}dk \left(\hat{a}^\dagger \hat{b}_{k}+\hat{b}^\dagger_{k}\hat{a}\right),
\end{split}
\label{hamiltonian}
\end{equation}
where $\omega_{s}$ is the resonance frequency of the KPO when no pump filed is applied, and $\hat{a}$ is the annihilation operator for the KPO.
The second and third terms represent the anharmonicity of the KPO and the effect of the pump~\cite{Goto2019S,Wang2019S}, respectively. 
$\beta$, $\omega_p$ and $\chi$ are the amplitude and angular frequency of the pump and the anharmonicity parameter of the KPO, respectively.
The fourth term is the Hamiltonian of the eigenmodes of the TL, and the fifth term is the interaction Hamiltonian between the KPO and TL.
Here, $\hat{b}_k$ is the annihilation operator of the mode with wave number $k$ in the TL; $v_b$ is the phase velocity of the microwave in the TL; $\kappa$ is the decay rate to the TL. 
We assume that the loss of microwave photons is negligible for simplicity.

In a frame rotating at $\omega_p/2$, the master equation for the KPO is represented as~\cite{Goto2019S}
\begin{equation}
\begin{split}
\frac{d\rho(t)}{dt}
&=
-i
\left[
\Delta \hat{a}^\dagger \hat{a}
-
\frac{\chi}{2}\hat{a}^\dagger \hat{a}^\dagger \hat{a}\hat{a}
+
\beta(\hat{a}^\dagger \hat{a}^\dagger+\hat{a}\hat{a})
,\rho(t)\right]
+\left[
\kappa_{ex}\hat{a}\rho(t)\hat{a}^{\dagger}
-
\frac{\kappa_{ex}}{2}\left\{\hat{a}^\dagger\hat{a},\rho(t)\right\}
\right]
,
\\
\end{split}
\label{master eq}
\end{equation}
where $\rho$ is the density operator, and $\Delta =\omega_s-\chi-\omega_p/2$.
The steady state of the master equation (\ref{master eq}) is approximated by the completely mixed state $(\ket{\alpha}\bra{\alpha}+\ket{-\alpha}\bra{-\alpha})/2$ of the coherent states $\ket{\alpha}$ and $\ket{-\alpha}$, where $\alpha$ is given by Eq.~(\ref{Puri}).

\section*{S2 Average of jump interval}
We obtain the average of the time interval between jumps, $E[T_{\rm i}]$, by using the binomial-coherent-state model. 
As explained in the main text, in this model, the KPO can only take either of $\ket{\pm \alpha}$ and jumps between the two states with a probability of $p=\Omega dt$ in time $dt$, where $\Omega$ is the rate of jumps. The average of the time interval between jumps is given by Eq.~(\ref{ET_4_4_22}).
Using Eq. (\ref{rho_4_5_22}), the expected value of $\hat{x}=(\hat{a}+\hat{a}^\dagger)/2$ is represented as

\begin{equation}
\begin{split}
\braket{\hat{x}}
&=
\sum_{k=2n}^{N}{}_N \mathrm{C}_k p^{k} (1-p)^{N-k} \mathrm{Tr}[\rho(t) \ket{\alpha}\bra{\alpha} ]
+
\sum_{k=2n+1}^{N}{}_N \mathrm{C}_k p^{k} (1-p)^{N-k} \mathrm{Tr}[\rho(t) \ket{-\alpha}\bra{-\alpha} ]
\\
&=
\sum_{k=2n}{}_N \mathrm{C}_k p^{k} (1-p)^{N-k} \mathrm{Re}[\alpha]
+
\sum_{k=2n+1}{}_N \mathrm{C}_k p^{k} (1-p)^{N-k}\mathrm{Re}[-\alpha]
\\
&=
\sum_{k=2n}{}_N \mathrm{C}_k p^{k} (1-p)^{N-k} \mathrm{Re}[\alpha]
-
\sum_{k=2n+1}{}_N \mathrm{C}_k p^{k} (1-p)^{N-k}\mathrm{Re}[\alpha]
\\
&=
\sum_{k=2n}{}_N \mathrm{C}_k (-1)^k p^{k} (1-p)^{N-k} \mathrm{Re}[\alpha]
+
\sum_{k=2n+1}{}_N \mathrm{C}_k (-1)^k p^{k} (1-p)^{N-k}\mathrm{Re}[\alpha]
\\
&=
\mathrm{Re}[\alpha]
\sum_k {}_N \mathrm{C}_k(-p)^k (1-p)^{N-k}
\\
&=
\mathrm{Re}[\alpha]
(-p+1-p)^N
\\
&=
\mathrm{Re}[\alpha]
(1-2p)^N,
\end{split}
\label{expect x}
\end{equation}

\noindent
where we used $N=t/dt$.
Taking the limit of $dt\rightarrow 0$, we obtain

\begin{equation}
\begin{split}
\lim_{dt\rightarrow 0}
\braket{\hat{x}}
&=
\lim_{dt\rightarrow 0}
\mathrm{Re}[\alpha]
(1-2p)^N
\\
&=
\lim_{dt\rightarrow 0}
\mathrm{Re}[\alpha]
(1-2\Omega dt)^{t/dt}
\\
&=
\mathrm{Re}[\alpha]
\exp(-2\Omega t)
.
\end{split}
\label{exp func}
\end{equation}

\begin{figure}[H]
  \centering
  \includegraphics[width=15cm]{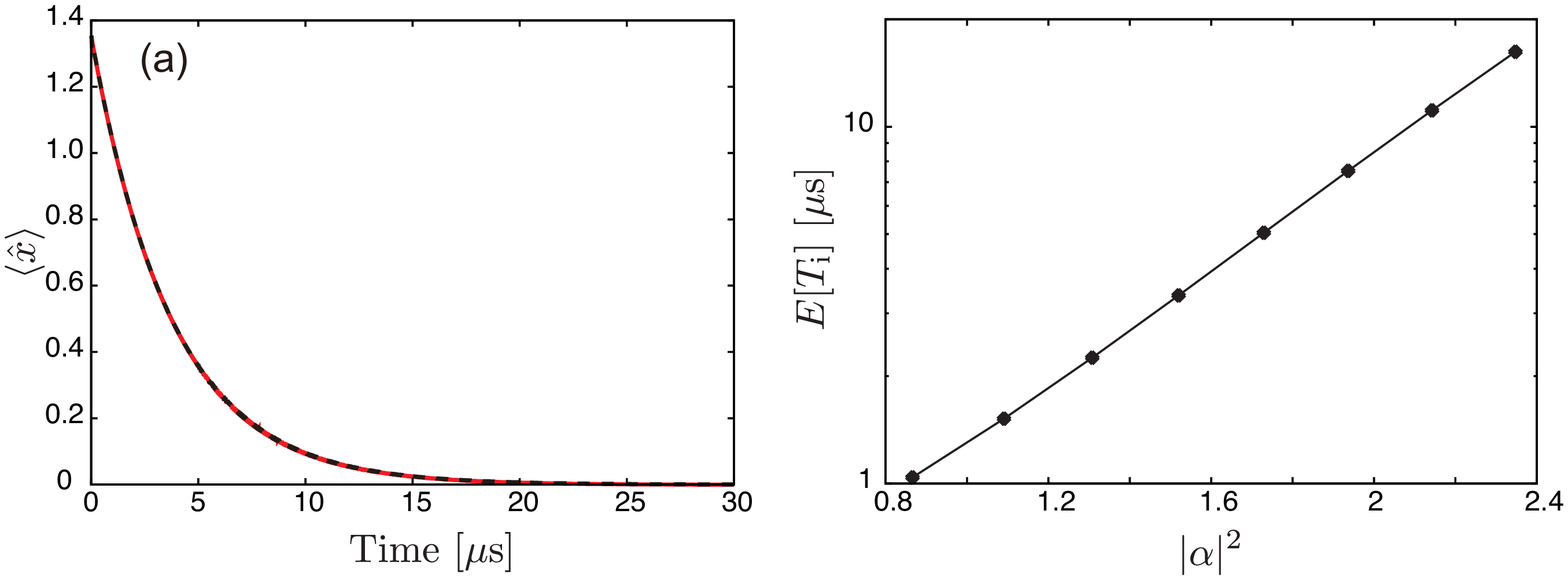}
  \caption{
(a): $\braket{\hat{x}}$ obtained by solving the master equation (\ref{master eq}) for $\alpha=1.38-0.18i$ (black dashed curve) and $\braket{\hat{x}}$ in Eq.~(\ref{exp func}) for $\Omega/2\pi=20~\rm{kHz}$ (red solid curve).
The two curves are almost overlapping.
(b): $E[T_{\rm i}]$ in Eq.~(\ref{ET_4_4_22}) with $\Omega$ obtained by the fitting of $\braket{\hat{x}}$ in Eq.~(\ref{exp func}) to numerical results of the master equation (\ref{master eq}).
($E[T_{\rm i}]$ is shown in a log scale.)
$\beta$ was changed to vary $|\alpha|$, while $\chi$ is fixed. The other parameters are the same as in Fig.~\ref{FidPM}.}.
\label{fitting2}
\end{figure}

\noindent
We can obtain $\Omega$ by fitting $\braket{\hat{x}}$ in Eq.~(\ref{exp func}) to the counterpart of the dynamics governed by the master equation~(\ref{master eq}).
(Note that $\rho(t)$ in Eq.~(\ref{rho_4_5_22}) coincides with the solution of the master equation in Eq.~(\ref{master eq}) when the binomial-coherent-state model is valid as explained in the main text.)
Figure \ref{fitting2}(a) presents the result of the fitting for $\alpha=1.38-0.18i$ as an example. The time dependence of $\braket{\hat{x}}$ in Eq.~(\ref{exp func}) with $\Omega/2\pi=20 \rm{kHz}$ matches well to the one obtained by solving the master equation (\ref{master eq}) for $\alpha=1.38-0.18i$. Figure \ref{fitting2}(b) shows $E[T_{\rm i}]$ as a function of $|\alpha|^2$. It is seen that $E[T_{\rm i}]$ exponentially increases with the increase of $|\alpha|^2$~~\cite{Puri2019S}.

\begin{thebibliography}{99}
\bibitem{Milburn1991} Milburn, G. J. \& Holmes, C. A. Quantum coherence and classical chaos in a pulsed parametric oscillator with a Kerr nonlinearity. \textit{Phys. Rev. A} \textbf{44,} 4704-4711 (1991).

\bibitem{Wielinga1993} Wielinga, B. \& G. J. Milburn, G. J. Quantum tunneling in a Kerr medium with parametric pumping. \textit{Phys. Rev. A} \textbf{48,} 2494-2496 (1993).

\bibitem{Goto2016} Goto, H. Bifurcation-based adiabatic quantum computation with a nonlinear oscillator network. \textit{Sci. Rep.} \textbf{6,} 21686 (2016).

\bibitem{Kirchmair2013}Kirchmair, G., Vlastakis, B., Leghtas, Z., Nigg, S. E., Paik, H., Ginossar, E., Mirrahimi, M., Frunzio, L., Girvin, S. M. \&  Schoelkopf, R. J. Observation of quantum state collapse and revival due to the single-photon Kerr effect. \textit{Nature} \textbf{495,} 205-209 (2013).

\bibitem{Goto2019} Goto, H. Quantum Computation Based on Quantum Adiabatic Bifurcations of Kerr-Nonlinear Parametric Oscillators. \textit{J. Phys. Soc. Jpn.} \textbf{88,} 061015 (2019).

\bibitem{Dykman2018} Dykman, M. I., Bruder, C., L\"{o}rch, N \& Zhang, Y. Interaction-induced time-symmetry breaking in driven quantum oscillators. \textit{Phys. Rev. B} \textbf{98,} 195444 (2018).

\bibitem{Rota2019}Rota, R., Minganti, F., Ciuti, C. \& Savona, V. Quantum Critical Regime in a Quadratically Driven Nonlinear Photonic Lattice. \textit{Phys. Rev. Lett.} \textbf{122,} 110405 (2019).

\bibitem{Meaney2014} Meaney, C. H., Nha, H., Duty, T. \& Milburn, G. J.  Quantum and classical nonlinear dynamics in a microwave cavity. \textit{EPJ Quantum Technol.} \textbf{1,} 7 (2014).

\bibitem{Wang2019} Wang, Z., Pechal, M., Wollack, E. A., Arrangoiz-Arriola, P., Gao, M., Lee, N. R. \& Safavi-Naeini, A. H. Quantum Dynamics of a Few-Photon Parametric Oscillator. \textit{Phys. Rev. X} \textbf{9,} 021049 (2019).

\bibitem{Grimm2020} Grimm, A., Frattini, N. E., Puri, S., Mundhada, S. O., 
Touzard, S., Mirrahimi, M., Girvin, S. M., Shankar, S. \& Devoret, M.
H.  Stabilization and operation of a Kerr-cat qubit. \textit{Nature} \textbf{584,} 205 (2020).

\bibitem{Tuckett2019}Tuckett, D. K., Darmawan, A. S., Chubb, C. T., Bravyi, S., Bartlett, S. D. \&  Flammia, S. T. Tailoring Surface Codes for Highly Biased Noise. \textit{Phys. Rev. X} \textbf{9,} 041031 (2019).

\bibitem{Ataides2021}Ataides, J. P. B., Tuckett, D. K., Bartlett, S. D., Flammia, S. T. \& Brown, B. J.  The XZZX surface code. \textit{Nat. Commun.} \textbf{12,} 2172 (2021).

\bibitem{Puri2017} Puri, S., Andersen, C. K., Grimsmo, A. L. \& Blais, A.
Quantum annealing with a network of all-to-all connected, two-photon driven Kerr nonlinear oscillators. \textit{Nat. Commun.} \textbf{8,} 15785 (2017).

\bibitem{Nigg2017} Nigg, S. E., L\"{o}rch, N. \& Tiwari, R. P. Robust quantum optimizer with full connectivity. \textit{Sci. Adv.} \textbf{3,} e1602273 (2017).

\bibitem{Zhao2018} Zhao, P., Jin, Z., Xu, P., Tan, X., Yu, H., \& Yu, Y., Two-Photon Driven Kerr Resonator for Quantum Annealing with Three-Dimensional Circuit QED. \textit{Phys. Rev. Applied} \textbf{10,} 024019 (2018).

\bibitem{Onodera2020} Onodera, T., Ng, E. \& McMahon, P. L. A quantum annealer with fully programmable all-to-all coupling via Floquet engineering. \textit{npj Quantum Inf.} \textbf{6,} 48 (2020).

\bibitem{Goto2020} Goto, H. \& Kanao, T. Quantum annealing using vacuum states as effective excited states of driven systems. \textit{Commun. Phys.} \textbf{3,} 235 (2020).

\bibitem{Kanao2021} Kanao, T and Goto, H. High-accuracy Ising machine using Kerr-nonlinear parametric oscillators with local four-body interactions. \textit{npj Quantum Inf.} \textbf{7,} 18 (2021).

\bibitem{Cochrane1999} Cochrane, P. T., Milburn, G. J. \& Munro, W. J. \textit{Phys. Rev. A} \textbf{59,} Macroscopically distinct quantum-superposition states as a bosonic code for amplitude damping. 2631-2634 (1999).

\bibitem{Puri2020} Puri, S. et al. Bias-preserving gates with stabilized cat qubits. \textit{Sci. Adv.} \textbf{6,} eaay5901 (2020).

\bibitem{Darmawan2021} Darmawan, A. S., Brown, B. J., Grimsmo, A. L.,  Tuckett, D. K. \& Puri, S. Practical Quantum Error Correction with the XZZX Code and Kerr-Cat Qubits. \textit{Phys. Rev. X} \textbf{2,} 030345 (2021).

\bibitem{Zhang2017} Zhang, Y. \& Dykman, M. I. Preparing quasienergy states on demand: A parametric oscillator. \textit{Phys. Rev. A} \textbf{95,} 053841 (2017).

\bibitem{Kanao2021-2} Kanao, T., Masuda, S., Kawabata, S. \& Goto, H. Quantum Gate for Kerr-Nonlinear Parametric Oscillator Using Effective Excited States. arXiv:2108.03091 (2021).

\bibitem{Xu2022}Xu, Q., Iverson, J. K., Brand\~{a}o, F. G. S. L. \& Jiang, L. Engineering fast bias-preserving gates on stabilized cat qubits. \textit{Phys. Rev. Research} \textbf{4,} 013082 (2022).

\bibitem{Kang2022} Kang, Y., Chen, Y., Wang, X., Song, J., Xia, Y., Miranowicz, A., Zheng, S. \& Nori, F. Nonadiabatic geometric quantum computation with cat qubits via invariant-based reverse engineering. \textit{Phys. Rev. Research} \textbf{4,} 013233 (2022).

\bibitem{Masuda2021} Masuda, S., Yamaguchi, A., Yamaji, T., Yamamoto, T., Ishikawa, T., Matsuzaki, Y. \& Kawabata, S. Theoretical study of reflection spectroscopy for superconducting quantum parametrons. \textit{New J. Phys.} \textbf{23,} 093023 (2021).

\bibitem{Yamaji2022} Yamaji, T., Kagami, S., Yamaguchi, A., Satoh, T.,  Koshino, K., Goto, H., Lin, Z. R., Nakamura, Y. \& Yamamoto, T. Spectroscopic observation of the crossover from a classical Duffing oscillator to a Kerr parametric oscillator. \textit{Phys. Rev. A} \textbf{105,} 023519 (2022).

\bibitem{Goto2018}Goto, H., Lin, Z. \& Nakamura, Y. Boltzmann sampling from the Ising model using quantum heating of coupled nonlinear oscillators. \textit{Sci. Rep.} \textbf{8,} 7154 (2018).

\bibitem{Masuda2021-2} Masuda, S., Ishikawa, T., Matsuzaki, Y. \& Kawabata, S. Controls of a superconducting quantum parametron under a strong pump field. \textit{Sci. Rep.} \textbf{11,} 11459 (2021).

\bibitem{Goto2021} Goto, H. \& Kanao, T. Chaos in coupled Kerr-nonlinear parametric oscillators. \textit{Phys. Rev. Research} \textbf{3,} 043196 (2021).

\bibitem{Minganti2016}Minganti, F., Bartolo, N., Lolli, J., Casteels, W. \& Ciuti, C. Exact results for Schr\"{o}dinger cats in driven-dissipative systems and their feedback control. \textit{Sci. Rep.} \textbf{6,} 26987 (2016).

\bibitem{Bartolo2017} Bartolo, N., Minganti, F., Lolli, J. \& Ciuti, C. Homodyne versus photon-counting quantum trajectories for dissipative Kerr resonators with two-photon driving. \textit{Eur. Phys. J. Spec. Top.} \textbf{226,} 2705 (2017).

\bibitem{Goto2016-2}Goto, H. Universal quantum computation with a nonlinear oscillator network. \textit{Phys. Rev. A} \textbf{93,} 050301 (2016).

\bibitem{Masuda2022}Masuda, S., Kanao, T., Goto, H., Matuszaki, Y., Ishikawa, T., \& Kawabata, S. Fast tunable coupling scheme of Kerr-nonlinear parametric oscillators based on shortcuts to adiabaticity. arXiv:2203.00226 (2022).

\bibitem{Wiseman2009} Wiseman, H. M. \& Milburn, G. J. \textit{Quantum Measurement and Control} (Cambridge University Press, 2009).

\bibitem{Puri2017-2} Puri, S., Boutin, S. \& Blais, A. Engineering the quantum states of light in a Kerr-nonlinear resonator by two-photon driving. \textit{npj Quantum Inf.} \textbf{3,} 18 (2017).

\bibitem{Nielsen2010} Nielsen, M. A.  \& Chuang, I. L. \textit{Quanyum Computation and Quantum information} (Cambridge University Press, Cambridge, UK, 10th anniversary ed., 2010).

\bibitem{QuTiP} Johansson, J. R., Nation, P. D. \& Nori, F. QuTiP 2: A Python framework for the dynamics of open quantum systems. \textit{Comp. Phys. Comm.} \textbf{184,} 1234 (2013).

\bibitem{Gambetta2008} Gambetta, J. \textit{et al.} Quantum trajectory approach to circuit QED: quantum jumps and Zeno effect. \textit{Phys. Rev. A} \textbf{77,} 012112(2008)

\bibitem{Puri2019}Puri, S. et al. Stabilized cat in a driven nonlinear cavity: a fault-tolerant error syndrome detector. \textit{Phys. Rev. X} \textbf{9,} 041009 (2019).

\bibitem{Jacob2006} Jacob, K. \& Steck, D. A. A Straightforward Introduction to Continuous Quantum Measurement. \textit{Contemp. Phys.} \textbf{47,} 279 (2006).
\end{thebibliography}

\begin{thebibliography}{99}
\bibitem{Goto2019S} Goto, H. Quantum Computation Based on Quantum Adiabatic Bifurcations of Kerr-Nonlinear Parametric Oscillators. \textit{J. Phys. Soc. Jpn.} \textbf{88,} 061015 (2019).
\bibitem{Wang2019S} Wang, Z., Pechal, M., Wollack, E. A., Arrangoiz-Arriola, P., Gao, M., Lee, N. R. \& Safavi-Naeini, A. H. Quantum Dynamics of a Few-Photon Parametric Oscillator. \textit{Phys. Rev. X} \textbf{9,} 021049 (2019).
\bibitem{Puri2019S}Puri, S. et al. Stabilized cat in a driven nonlinear cavity: a fault-tolerant error syndrome detector. \textit{Phys. Rev. X} \textbf{9,} 041009 (2019).
\end{thebibliography}
\end{document}